\documentclass[12pt, twoside]{article} 
\usepackage{amssymb}
\usepackage{amsmath} 
\usepackage{amscd} 
\usepackage{amsfonts} 
\usepackage{amsthm} 
\usepackage{eucal}
\usepackage{latexsym} 
\theoremstyle{definition} 
\newtheorem{defn}{Definition}[section] 
\theoremstyle{plain} 
\newtheorem{thm}{Theorem}[section] 
\newtheorem{lem}{Lemma}[section] 
\theoremstyle{remark} 
\newtheorem{rem}{Remark}[section] 
\newtheorem{ax}{Axiom}
\newcommand{\field}[1]{\mathbb{#1}} 
\newcommand{\F}{\field{F}} 
\newcommand{\R}{\field{R}} 
\newcommand{\C}{\field{C}}
\newcommand{\N}{\field{N}} 
\newcommand{\HH}{\field{H}}
\newcommand{\bo}[1]{\boldsymbol{#1}} 
\newcommand{\mf}[1]{\mathsf{#1}} 
\newcommand{\mc}[1]{\mathcal{#1}} 

\newcommand{\mk}[1]{\mathfrak{#1}} 
\begin{document} 
\title{Hyper-Hamiltonian quantum mechanics} 
\author{Vladimir Trifonov \\ trifonov@member.ams.org} 
\date{}\maketitle 
\begin{abstract} 
We present a modification of quantum mechanics with a \emph{possible} \emph{worlds} semantics. It is shown that `gauge' degrees of freedom along possible worlds can be used to encode gravitational information. 
\end{abstract} 
\section{Preliminaries.} 
It is known that Schr\"{o}dinger evolution is Hamiltonian evolution on a manifold \emph{canonically} \emph{associated} to the Hilbert space (see, e.~g., \cite{Sch96} and references therein). In this note we combine the kinematics of the semi-geometrical approach with the hyper-Hamiltonian formalism (\cite{GM02}, \cite{MT02}) to incorporate gravity into the description of quantum systems.  
\begin{rem}[Notations] $\N$ denotes natural numbers, $\R$ and $\C$ are the fields of real and complex numbers, respectively. Small Latin indices $i, j, k$ take values in $\N$; small Greek indices, $\alpha, \beta, \gamma$ and small Latin indices $p, q$ \emph{always} run $0$ to $3$ and $1$ to $3$, respectively. Summation is assumed on repeated indices of different levels unless explicitly indicated otherwise. For structural clarity we use $\Box$ at the end of a \emph{Proof}, and each  \emph{Remark} ends with the sign appearing at the end of this line. \hfill $\Diamond$ \end{rem} 
\begin{defn} An $\F$-\emph{algebra}, $A$, is an ordered pair $(S_A, \bo{\mf{A}})$, where $S_A$ is a vector space  over a field $\F$, and $\bo{\mf{A}}$ is a $(\begin{smallmatrix} 1\\2 \end{smallmatrix})$-tensor  on $S_A$, called the \emph{structure} \emph{tensor} of $A$. Each vector $\bo{a}$ of $S_A$ is called an \emph{element} of $A$, denoted $\bo{a} \in A$.  The \emph{dimensionality} of $A$ is that of $S_A$. \end{defn} 
\begin{rem} The tensor $\bo{\mf{A}}$ induces a binary operation $S_A \times S_A \to S_A$, called the \emph{multiplication} of $A$: to each pair of vectors $(\bo{a}, \bo{b})$ the tensor $\bo{\mf{A}}$ associates a vector $\bo{ab} : S^*_A \to \F$, such that $\bo{ab}(\tilde{\bo{\tau}}) =  \bo{\mf{A}}(\tilde{\bo{\tau}}, \bo{a}, \bo{b}), \forall \bo{\tau} \in S^*_A$. An $\F$-algebra with an associative multiplication is called \emph{associative}. An element $\bo{\imath}$, such that $\bo{a\imath} = \bo{\imath a} =\bo{a}, \forall \bo{a} \in A$ is called an \emph{identity} of $A$.  \hfill $\Diamond$ \end{rem} 
\begin{defn} For each $\F$-algebra $A = (S_A, \bo{\mf{A}})$, an $\F$-algebra $[A] = (S_A, \bar{\bo{\mf{A}}})$, where $\bar{\bo{\mf{A}}}(\tilde{\bo{\tau}}, \bo{a}, \bo{b}) := \bo{\mf{A}}(\tilde{\bo{\tau}}, \bo{a}, \bo{b}) - \bo{\mf{A}}(\tilde{\bo{\tau}}, \bo{b}, \bo{a}), \forall \bo{a}, \bo{b} \in S_A$, is called the \emph{commutator} algebra of $A$. \end{defn}
\begin{defn} For an $\F$-algebra $A$ and each nonzero one-form $\tilde{\bo{\tau}} \in S^*_A$, the tensor $\bo{\mf{G}}^A := \tilde{\bo{\tau}} \lrcorner \bo{\mf{A}}$ is called a \emph{principal} \emph{metric} of $A$, just in case it is symmetric, where $\lrcorner$ denotes contraction on the first index. \end{defn}
\begin{rem} \label{VECMAN} For an $\R$-algebra $A$, its vector space $S_A$ canonically generates a (linear) manifold $\mc{S}_A$ with the same carrier, so we have a bijection  $\mc{J} :\mc{S}_A \to S_A$. We distinguish between the two objects, using the normal ($a, u, ...$) and bold ($\bo{a}, \bo{u}, ...$) fonts, respectively,  to denote their elements. The tangent space $T_a\mc{S}_A$ is identified with $S_A$ at each point $a \in \mc{S}_A$ via an isomorphism $\mc{J}^*_a : T_a\mc{S}_A \to S_A$ sending a tangent vector to the curve $\mu : \R \to \mc{S}_A, \mu(t) =  a + t\bo{u}$, at the point $\mu(0)$, to the vector $\bo{u} \in S_A$, with the ``total'' map $\mc{J}^* : T\mc{S}_A \to S_A$. A linear map $\bo{F} : S_A \to S_A$ induces a vector field $\bo{f} : \mc{S}_A \to T\mc{S}_A$ on $\mc{S}_A$, such that the following diagram commutes, 
\begin{equation} \label{VM} \begin{CD} \mc{S}_A @> \bo{f} >> T\mc{S}_A \\ @V \mc{J} VV @VV \mc{J}^* V\\ 
S_A @> \bo{F} >> S_A \end{CD} \quad . \end{equation} 
\hfill $\Diamond$ \end{rem}
\begin{defn} A left module over an algebra $A$ is a vector space $V$ together with a bilinear map $A \times V \to V$. \end{defn}
\begin{defn} A finite dimensional associative $\R$-algebra with an identity is called a \emph{unital} algebra. \end{defn}
\begin{lem} The set $\mc{A}$ of all invertible elements of a unital algebra $A$ is a submanifold of $\mc{S}_A$, and a Lie group with respect to the multiplication of $A$. The Lie algebra of $\mc{A}$ is the commutator algebra $[A]$. The vector space of the Lie algebra of $\mc{A}$ is $S_A$, and therefore $S_A$ is the tangent space $T_{\bo{\imath}}\mc{A}$ of $\mc{A}$ at the identity of the group $\mc{A}$. \end{lem} 
\begin{proof} See, for example, \cite{Pos82} for a proof of this simple lemma. \end{proof} 
\begin{rem} For each basis $(\bo{e}_j)$ on the vector space $S_A$ of a unital algebra, there is a natural basis (field) on $\mc{A}$, namely the basis $(\hat{\bo{e}}_j)$ of left invariant vector fields generated by $(\bo{e}_j)$. We call $(\hat{\bo{e}}_j)$ a \emph{proper} basis. \hfill $\Diamond$ \end{rem}
\begin{defn} For a unital algebra $A$, let $(\hat{\bo{e}}_j)$ be a  proper basis on $\mc{A}$, generated by a basis $(\bo{e}_j)$ on $S_A$. The \emph{structure} \emph{field} of the Lie group $\mc{A}$ is a tensor field $\bo{\mc{A}}$ on $\mc{A}$, assigning to each point $a \in \mc{A}$ a $(\begin{smallmatrix} 1\\2 \end{smallmatrix})$-tensor $\bo{\mc{A}}_a$ on $T_a\mc{A}$, with components $(\mc{A}_a)^i_{jk}$ in the basis $(\hat{\bo{e}}_j)$, defined by 
\begin{displaymath} (\mc{A}_a)^i_{jk} := \mf{A}^i_{jk} , \quad \forall a \in \mc{A} ,  \end{displaymath} where $\mf{A}^i_{jk}$ are the components of the structure tensor $\bo{\mf{A}}$ in the basis $(\bo{e}_j)$. \end{defn} 
\begin{defn} For a unital algebra $A$ and a one-form field $\tilde{\bo{\tau}}$ on $\mc{A}$,  a tensor field $\bo{\mf{g}}^{\mc{A}} = \tilde{\bo{\tau}} \lrcorner \bo{\mc{A}}$   is called a \emph{proper} \emph{metric} (field) of $\mc{A}$, provided it is symmetric. \end{defn}
\section{Quaternion algebra} \begin{defn} A four dimensional $\R$-algebra, $\HH = (S_{\HH}, \bo{\mf{H}})$, is called a \emph{quaternion} \emph{algebra} (with \emph{quaternions} as its elements), if there is a basis $(\bo{e}_{\beta})$ on $S_{\HH}$, in which the components of the structure tensor $\bo{\mf{H}}$ are given by the entries of the following matrices, 
\begin{multline} \label{QST} \mf{H}^0_{\alpha \beta} = 
\begin{pmatrix} 1&0&0&0\\0&-1&0&0\\0&0&-1&0\\0&0&0&-1 \end{pmatrix},\ 
\mf{H}^1_{\alpha \beta} = \begin{pmatrix} 0&1&0&0\\1&0&0&0\\0&0&0&1\\0&0&-
1&0 \end{pmatrix}, \\ \mf{H}^2_{\alpha \beta} = \begin{pmatrix} 
0&0&1&0\\0&0&0&-1\\1&0&0&0\\0&1&0&0 \end{pmatrix},\ \mf{H}^3_{\alpha 
\beta} = \begin{pmatrix} 0&0&0&1\\0&0&1&0\\0&-1&0&0\\1&0&0&0 
\end{pmatrix}. \end{multline} \end{defn} 
Such a basis is called a \emph{canonical} \emph{basis}, and its vectors are denoted $\bo{1}$, $\bo{i}$, $\bo{j}$, $\bo{k}$. A quaternion algebra is unital, with the first vector  of the canonical basis, $\bo{1}$, as its identity. Since $(\bo{1}$, $\bo{i}$, $\bo{j}$, $\bo{k})$ is a basis on a real vector space, any quaternion $\bo{a}$ can be presented as $a^0\bo{1} + a^1\bo{i} + a^2\bo{j} + a^3\bo{k}, a^{\beta} \in \R$. A quaternion $\bar{\bo{a}} = a^0\bo{1} - a^1\bo{i} - a^2\bo{j} - a^3\bo{k}$ is called \emph{conjugate} to $\bo{a}$. We refer to $a^0$ and $a^p\bo{i}_p$ as the \emph{real} and  \emph{imaginary} \emph{part} of $\bo{a}$, respectively. Quaternions of the form $a^0\bo{1}$ are in one-to-one correspondence with real numbers, which is often denoted, with certain notational abuse, as $\R \subset \HH$. 
\begin{rem} A linear transformation $S_{\HH} \to S_{\HH}$ with the following components in the canonical basis, 
\begin{displaymath} \begin{pmatrix} 1 & 0 \\ 0& \bo{\mf{B}} \end{pmatrix}, \bo{\mf{B}} \in SO(3), \end{displaymath} 
takes $(\bo{1}$, $\bo{i}$, $\bo{j}$, $\bo{k})$ to a basis $(\bo{i}_{\beta})$ in which the components \eqref{QST} of the structure tensor will \emph{not} change, and neither will the multiplicative behavior of vectors of $(\bo{i}_{\beta})$. Thus, we have a class of canonical bases parameterized by elements of $SO(3)$. The notions of conjugation as well as the real and imaginary parts are clearly invariant under the canonical basis change. \hfill $\Diamond$ \end{rem} 
\begin{rem} The vector space $S_{\HH}$ induces, via the process \eqref{VM}, a manifold $\mc{S}_{\HH}$, with the canonical identification $T_a\mc{S}_{\HH} \cong S_{\HH}, \forall a \in \mc{S}_{\HH}$. \hfill $\Diamond$ \end{rem} 
\section{Principal metrics of $\HH$.} 
\begin{lem} Each principal metric of a quaternion algebra is a Minkowski metric. \end{lem} \begin{proof} For a basis $\bo{e}_{\beta}$ in $S_{\HH}$ let 
$\tilde{\tau}_{\beta}$ be the components of a one-form $\tilde{\bo{\tau}}$ in the dual basis $\bo{e}^{\beta}$.  Then the components $\mf{G}^{\HH}_{\alpha \beta}$ of the principal metric are 
\begin{displaymath} \mf{G}^{\HH}_{\alpha \beta} =  \tilde{\tau}_{\gamma}\mf{H}^{\gamma}_{\alpha \beta}, \end{displaymath} 
where $\mf{H}^{\gamma}_{\alpha\beta}$ are the components of $\bo{\mf{H}}$. In a canonical   basis we have: 
\begin{displaymath} \mf{G}^{\HH}_{\alpha \beta} = \begin{pmatrix} 
\tilde{\tau}_0& \tilde{\tau}_1& \tilde{\tau}_2& \tilde{\tau}_3\\ 
\tilde{\tau}_1&-\tilde{\tau}_0& \tilde{\tau}_3&-\tilde{\tau}_2\\ 
\tilde{\tau}_2&-\tilde{\tau}_3&-\tilde{\tau}_0& \tilde{\tau}_1\\ 
\tilde{\tau}_3& \tilde{\tau}_2&-\tilde{\tau}_1&-\tilde{\tau}_0 \end{pmatrix}. \end{displaymath} 
The only way to make this matrix symmetric is to put $\tilde{\tau}_1=-\tilde{\tau}_1$, $\tilde{\tau}_2=-\tilde{\tau}_2$,  $\tilde{\tau}_3=-\tilde{\tau}_3$, which yields $\tilde{\tau}_1=\tilde{\tau}_2=\tilde{\tau}_3=0$: 
\begin{displaymath} \mf{G}^{\HH}_{\alpha \beta} = \begin{pmatrix} 
\tilde{\tau}_0& 0& 0& 0\\ 
0&-\tilde{\tau}_0& 0&0\\ 
0&0&-\tilde{\tau}_0& 0\\ 
0& 0&0&-\tilde{\tau}_0 \end{pmatrix}. \end{displaymath} 
Thus for each $\tilde{\tau} \in \R \setminus \{0\}$, the quaternion algebra $\HH$ has a principal metric of signature 2, generated by a one-form $\tilde{\bo{\tau}}$ with the components $(\tilde{\tau}, 0, 0, 0)$ in the canonical basis, which concludes the proof, \cite{Tri95}. \end{proof} 
\section{Natural coordinates and bases on $\HH$ and $\mc{H}$.} 
Every nonzero element of a quaternion algebra, $\HH$, is invertible. Therefore the Lie group $\mc{H}$ of invertible elements of $\HH$ is the set of nonzero quaternions together with the multiplication on $\HH$.  
Each canonical basis $(\bo{i}_{\beta})$ induces a  coordinate system $(w$, $x$, $y$, $z)$ on $\mc{S}_{\HH}$  and therefore also on its submanifold $\mc{H}$: a quaternion $a = a^{\beta}\bo{i}_{\beta}$ is assigned coordinates $(w = a^0$, $x = a^1$, $y = a^2$, $z = a^3)$. This coordinate system covers both $\mc{S}_{\HH}$ and $\mc{H}$ with a single patch.  Since $0 \notin \mc{H}$, at least one of the coordinates is always nonzero for any point $a \in \mc{H}$. For a differentiable function $R : \R \to \R\setminus\{0\}$ there is a system of natural spherical coordinates  $(\eta$, $\chi$, $\theta$, $\phi)$ on $\mc{H}$, related to the canonical coordinates by 
\begin{multline*} 
w = R(\eta)\cos(\chi), \quad x = R(\eta)\sin(\chi)\sin(\theta)\cos(\phi), \\
y = R(\eta)\sin(\chi)\sin(\theta)\sin(\phi), \quad z = R(\eta)\sin(\chi)\cos(\theta) . 
\end{multline*}
There are several natural basis fields on 
$\mc{H}$ induced by each canonical basis $(\bo{i}_{\beta})$ on $S_{\HH}$. First of all, we   have a (noncoordinate) proper basis $(\bo{\hat{\imath}}_{\beta})$. There are also two coordinate bases, $(\partial_w$, $\partial_x$, $\partial_y$, $\partial_z)$ and the corresponding spherical coordinate basis $(\partial_{\eta}$, $\partial_{\chi}$, $\partial_{\theta}$, $\partial_{\phi})$.  
\section{Proper metrics of $\mc{H}$.}
\begin{thm} All proper metrics of $\mc{H}$ are closed  Friedmann-Lema\^{\i}tre-Robertson-Walker. \end{thm}  
\begin{proof}  A left invariant vector field $\bo{f}$ on $\mc{H}$, generated by a vector $\bo{f}_0 \in S_{\HH}$, with components $(f_0^{\alpha})$ in the coordinate basis, associates to each point $a \in \mc{H}$ with coordinates $(w$, $x$, $y$, $z)$ a vector $\bo{f}(a) \in T_a\mc{H}$ with the components $f^{\beta} = (a\bo{f}_0)^{\beta}$ in the coordinate basis: 
\begin{multline} \label{LVFIELDS} f^0 = wf_0^0 - xf_0^1 - yf_0^2 - zf_0^3 , \quad f^1 = wf_0^1 + xf_0^0 + yf_0^3 - zf_0^2 , \\ f^2 = wf_0^2 - xf_0^3 + yf_0^0 + zf_0^1 , 
\quad f^3 = wf_0^3 + xf_0^2 - yf_0^1 + zf_0^0 . \end{multline} 
The system \eqref{LVFIELDS} contains sufficient information to compute transformation between the bases. For example, the transformation between the spherical and the proper basis is given by 
\begin{displaymath} \begin{pmatrix} 
R/\dot{R} & 0 & 0 & 0\\ 
0 & \sin{\theta}\cos{\phi} & \sin{\theta}\sin{\phi} & \cos{\theta} \\ 
0 & \frac{\cos{\chi}\cos{\theta}\cos{\phi}+\sin{\chi}\sin{\phi}}{\sin{\chi}} & 
\frac{\cos{\chi}\cos{\theta}\sin{\phi}+\sin{\chi}\cos{\phi}}{\sin{\chi}} & 
\frac{\cos{\chi}\sin{\theta}}{\sin{\chi}}\\ 
0 & \frac{\sin{\chi}\cos{\theta}\cos{\phi}-\cos{\chi}\sin{\phi}}{\sin{\chi}\sin{\theta}} & 
\frac{\sin{\chi}\cos{\theta}\sin{\phi}+\cos{\chi}\cos{\phi}}{\sin{\chi}\sin{\theta}} & -1 
\end{pmatrix},\end{displaymath} 
where $\dot{R} \not\equiv 0$. We know that for each $\tilde{\tau} \in \R$ a one-form $\tilde{\bo{\tau}}$ with components $(\tilde{\tau}$, $0$, $0$, $0)$ in (the dual of) a canonical  basis $(\bo{i}_{\beta})$ converts the structure tensor into a principal metric. In order for a one-form field $\tilde{\bo{\tau}}$ to generate a proper metric, it has to have the components $(\tilde{\tau}(a)$, $0$, $0$, $0)$ in the proper basis $(\bo{\hat{\imath}}_{\beta})$, with $\tilde{\tau}(a) : \mc{H} \to \R$. It is easy to see that only a one-form of this kind is capable of producing a symmetrical tensor. In the spherical basis $\tilde{\tau}$ has components $((R/\dot{R})\tilde{\tau}$, $0$, $0$, $0)$, which means that $\tilde{\tau}$ must depend only on $\eta$. Then its contraction with the structure field $\bo{\mc{H}}$ produces a proper metric whose components in the spherical basis are 
\begin{displaymath} g^{\mc{H}}_{\alpha \beta} = \begin{pmatrix} 
\tilde{\tau}(\eta)(\frac{\dot{R}}{R})^2&0&0&0\\0&-\tilde{\tau}(\eta)&0&0\\ 
0&0&-\tilde{\tau}(\eta){\sin^2(\chi)} &0\\ 
0&0&0&-\tilde{\tau}(\eta){\sin^2(\chi)} {\sin^2(\theta)} \end{pmatrix}. \end{displaymath} 
If $\tilde{\tau}(\eta) > 0$, we take $R(\eta)$ such that $\tilde{\tau}(\eta)(\frac{\dot{R}}{R})^2 = 
1$, which yields \begin{equation} \label{+R} R(\eta) = 
exp{\int\frac{d\eta}{\genfrac{}{}{0pt}{3}{+}{-}\sqrt{\tilde{\tau}(\eta)}}} ; \quad 
\tilde{\tau}(\eta) > 0 . \end{equation} 
In other words, with $R(\eta)$ satisfying \eqref{+R}, the metric is 
closed Friedmann-Lema\^{\i}tre-Robertson-Walker: \begin{displaymath} 
g^{\mc{H}}_{\alpha \beta} = \begin{pmatrix} 
1&0&0&0\\ 0&-\mf{a}^2&0&0\\ 0&0&-\mf{a}^2{\sin^2(\chi)} &0\\ 
0&0&0&-\mf{a}^2{\sin^2(\chi)} {\sin^2(\theta)} \end{pmatrix}, \end{displaymath} 
where the ``expansion factor'' $\mf{a}(\eta) := {\sqrt{\tilde{\tau}(\eta)}}$. The case  
$\tilde{\tau}(\eta) < 0$ is treated similarly, with $\mf{a}(\eta) := {\sqrt{-\tilde{\tau}(\eta)}}$, \cite{Tri03}.  \end{proof}
\section{Hyper-Hamiltonian dynamics.} 
\begin{defn} Let $\mc{M}$ and $\mc{N}$ be hyperk\"{a}hler manifolds with hypercomplex structures $\mc{I}^{\mc{M}}_p$ and $\mc{I}^{\mc{N}}_p$, respectively. A smooth map $\mk{F} : \mc{M} \to \mc{N}$ is called \emph{quaternionic} if there exists an $SO(3)$ matrix $\bo{\mf{B}}$ such that 
\begin{displaymath} \sum_{p,q=1}^3\mf{B}_{pq}\mc{I}^{\mc{N}}_q \circ d\mk{F} \circ \mc{I}^{\mc{M}}_p = d\mk{F} , \end{displaymath} where $d\mk{F}$ is the differential of $\mk{F}$, \cite{CL00}. \end{defn} 
For a canonical basis $(\bo{i}_{\beta})$ on $\HH$ we can express the values of a quaternionic map $\mk{F} : \mc{M} \to \mc{S}_{\HH}$ in the canonical coordinates $(w, x, y, z)$, which produces a quadruple $(\mk{F}^{\beta})$ of real-valued functions on $\mc{M}$,  the \emph{components} of $\mk{F}$ \emph{in} the basis $(\bo{i}_{\beta})$. 
\begin{defn} For a quaternionic map $\mk{F} : \mc{M} \to \mc{S}_{\HH}$, its  \emph{hyper}-\emph{Hamiltonian} vector field $\bo{f}$  on $\mc{M}$ is defined by 
\begin{equation} \label{HHE1} d\mk{F}^0 = \bo{f} \lrcorner \bo{g} , \end{equation} 
where $\bo{g}$ is the Riemannian metric on $\mc{M}$. \end{defn} 
\begin{rem} Since the decomposition of $\mk{F}$ into real and imaginary parts is invariant under the canonical basis change, so is the definition of a hyper-Hamiltonian vector field. \hfill $\Diamond$ \end{rem} 
The notion of a hyper-Hamiltonian vector field was introduced  in \cite{GM02} and \cite{MT02}, where it was shown that \eqref{HHE1} is a superposition of three Hamiltonian evolutions: 
\begin{displaymath} \label{HHE2} \bo{f} = \bo{f}_1 + \bo{f}_2 + \bo{f}_3 , \quad d\mk{F}^p = \bo{f}_p \lrcorner \tilde{\bo{\omega}}^p , \end{displaymath} 
where $(\tilde{\bo{\omega}}^p)$ is the triple  of symplectic forms on $\mc{M}$, and there is \emph{no} \emph{summation} on $p$. An example of an integrable hyper-Hamiltonian system (\emph{quaternionic} \emph{oscillator}) is given in \cite{GM02}. 

\section{Kinematics.} \label{KINEMATICS}
\begin{defn} A left module $V$ over $\HH$ is called a \emph{quaternionic} \emph{vector} \emph{space}. \end{defn} 
\begin{defn} A map $\widehat{\bo{F}} : V \to V$ is called a \emph{quaternion} \emph{linear} \emph{operator}, if 
\begin{displaymath} \widehat{\bo{F}}(\bo{a}\bo{\phi}) = \bo{a}\widehat{\bo{F}}(\bo{\phi}), \quad \forall \bo{\phi} \in V, \quad \forall \bo{a} \in \HH . \end{displaymath} \end{defn} 
\begin{defn} A \emph{quaternionic}  \emph{Hilbert} \emph{space} is an ordered pair $(V, \langle\cdot\mid\cdot\rangle)$, where $V$ is a quaternionic vector space,  and $\langle\cdot\mid\cdot\rangle$ is a map $V \times V \to \HH$, called a \emph{Hermitian} \emph{inner} \emph{product}, such that  
\begin{multline*} \quad \langle\bo{\phi} \mid \bo{\psi} + \bo{\xi}\rangle = \langle\bo{\phi}, \bo{\psi}\rangle + \langle\bo{\phi} \mid \bo{\xi}\rangle, \\ \langle\bo{\phi} \mid \bo{\psi}\rangle = \overline{\langle\bo{\psi} \mid \bo{\phi}\rangle}, \quad \langle\bo{\phi}, \bo{a}\bo{\psi}\rangle = \bo{a}\langle\bo{\phi}, \bo{\psi}\rangle , \\ 
\parallel\phi\parallel^2 := \langle\bo{\phi}, \bo{\phi}\rangle \in \R,  \parallel\bo{\phi}\parallel^2 > 0, \quad \forall \bo{\phi} \neq \bo{0}, \\ \forall \bo{\phi}, \bo{\psi}, \bo{\xi} \in V , \forall \bo{a} \in \HH , \end{multline*} 
and the diagonal $\parallel\cdot\parallel$ induces a topology on $V$, relative to which $V$ is \emph{separable} and \emph{complete} \cite{Ad95}. \end{defn} 
\begin{defn} For a quaternionic Hilbert space $(V, \langle\cdot\mid\cdot\rangle)$ and a quaternion linear operator $\widehat{\bo{F}}$ on $V$, its \emph{adjoint} (\emph{with} \emph{respect} \emph{to}  $\langle\cdot\mid\cdot\rangle$) is a quaternion linear operator $\widehat{\bo{F}}^{\dagger}$ on $V$, such that 
\begin{displaymath} \langle\bo{\phi} \mid \widehat{\bo{F}}(\bo{\psi})\rangle = \langle\widehat{\bo{F}}^{\dagger}(\bo{\phi}) \mid \bo{\psi}\rangle , \quad \forall \bo{\phi}, \bo{\psi} \in V . \end{displaymath} 
$\widehat{\bo{F}}$ is called \emph{(anti-)Hermitian} if it coincides with (the negative of) its adjoint.  \end{defn}
Given a canonical basis $\bo{i}_{\beta}$, an $n$ dimensional quaternionic Hilbert space $V$ generates, via the process \eqref{VM} a real $4n$ dimensional vector space, which we denote  $S_V$, and the latter induces a real $4n$ dimensional manifold $\mc{S}_V$. A quaternion linear operator $\widehat{\bo{F}} : V \to V$ induces a linear operator $\bo{F}$ on $S_V$, $\bo{F}(\bo{\phi}) := -\widehat{\bo{F}}(\bo{\phi})$, which in turn generates the vector field $\bo{f} : \mc{S}_V \to T\mc{S}_V$.
At each point $\xi \in \mc{S}_V$ the Hermitian inner product generates a map $\xi^{\prime} : T_{\xi}\mc{S}_V \times T_{\xi}\mc{S}_V \to \HH$, whose values can be  decomposed in the basis $\bo{i}_{\beta}$:
\begin{displaymath} \xi^{\prime}(\bo{\phi}, \bo{\psi}) = \frac{1}{2}\bo{g}_{\xi}(\bo{\phi}, \bo{\psi})\bo{i}_0 + \frac{1}{2}\tilde{\bo{\omega}}^p_{\xi}(\bo{\phi}, \bo{\psi})\bo{i}_p , \quad  \forall \bo{\phi}, \bo{\psi} \in T_{\xi}\mc{S}_V. \end{displaymath} 
where $\bo{g}_{\xi}$ is a positive-definite inner product and $(\tilde{\bo{\omega}}^p_{\xi})$ is a triple of two-forms on each $T_{\xi}\mc{S}_V$. The properties of the Hermitian inner product ensure that the maps $\bo{g}_{\xi}$ and $(\tilde{\bo{\omega}}^p_{\xi})$ constitute strongly nondegenerate tensor fields on  $\mc{S}_V$, namely, a Riemannian metric $\bo{g}$ and a triple of  symplectic forms, $(\tilde{\bo{\omega}}^p)$. In the same manner, a quadruple of operators $\bo{\imath}^{\prime}_{\beta}$ on $V$, defined by 
\begin{displaymath} \bo{\imath}^{\prime}_{\beta}(\bo{\phi}) := \bo{i}_{\beta}\bo{\phi}, \quad \forall \bo{\phi} \in V \end{displaymath} 
generate a quadruple $(\mc{I}_{\beta})$ of $(\begin{smallmatrix} 1 \\ 1 \end{smallmatrix})$-tensor fields on $\mc{S}_V$. The first one, $\mc{I}_0$ is the identity transformation on each tangent space $T_{\xi}\mc{S}_V$, and the last three make up an almost hypercomplex structure on $\mc{S}_V$. These tensors form a hyperk\"{a}hler structure on $\mc{S}_V$:  
\begin{multline*} \bo{g}(\bo{\phi}, \bo{\psi}) = \bo{g}(\mc{I}_p(\bo{\phi}), \mc{I}_p(\bo{\psi})), \\ \bo{g}(\bo{\phi}, \bo{\psi}) = \tilde{\bo{\omega}}^p(\bo{\phi}, \mc{I}_p(\bo{\psi})), \\ \tilde{\bo{\omega}}^p(\bo{\phi}, \bo{\psi}) = \tilde{\bo{\omega}}^p(\mc{I}_p(\bo{\phi}), \mc{I}_p(\bo{\psi})) , \end{multline*} (no summation on $p$), which makes $(\mc{I}_p)$ a hypercomplex structure on $\mc{S}_V$. 
\section{Axioms of HH quantum mechanics.}
Since $V$ is a left module over $\HH$, there is a left action of the group $\mc{H}$ on the set of nonzero vectors of $V$, which induces a natural principal $\mc{H}$-bundle $(\mc{P}, \mc{P}/\HH, \pi, \mc{H})$, with the total space $\mc{P} := \mc{S}_V \setminus \{0\}$. 
The base space is the quaternionic projective space $\mc{P}/\HH$, and the projection $\pi$ assigns to each $\phi$ its orbit $\mc{W}_{\phi}$.  
The standard fiber is $\mc{H}$. The fiber diffeomorphism $\sigma : \mc{H} \cong \mc{W}_{\phi}$ assigns to each $\bo{a} \in \mc{H}$ a point $\bo{a}\phi \in \mc{W}_{\phi}$. 
\par In order to avoid technical complications, the axioms below are given for the case of discrete set of eigenvalues with no degeneracy.  
\begin{ax}[The system and the perception template]  To a physical system the observer associates a bundle $\mc{P}$, called the \emph{propensity} \emph{domain}, together with a smooth real function $\mc{H} \to \R$, such that $\bo{\mf{g}}^{\mc{H}} := d\tau \lrcorner \bo{\mc{H}}$ is symmetric. The system is characterized by its \emph{local} \emph{propensities}, ordered triples $(\bo{\phi}, \mc{W}_{\phi}, \phi)$, where $\bo{\phi} \in V$ is called a \emph{state} of the system, and $\phi \in \mc{P}$ is a \emph{presence} \emph{mode} of the state $\bo{\phi}$ in a \emph{possible} \emph{world} $\mc{W}_{\phi}$. 
The manifold $\mc{H}$ is interpreted as the \emph{perception} \emph{template} (of the observer \emph{with} \emph{respect} to the system), with $\tau$ and $\bo{\mf{g}}^{\mc{H}}$ as its  \emph{global} \emph{time} and \emph{proper} \emph{metric}, respectively. \end{ax}
\begin{ax}[Observables] Each measurable quantity is represented by an ordered triple $(\mk{F}, \widehat{\bo{F}}, \bo{f})$, where $\mk{F}$ is a quaternionic map $\mc{P} \to \mc{S}_{\HH}$, called an \emph{observable}, $\widehat{\bo{F}}$ is an anti-Hermitian operator $V \to V$, called the \emph{lineal} of the observable, such that expectation of $\widehat{\bo{F}}$ coincides with the imaginary part of $\mk{F}$; and $\bo{f}$ is a vector field on $\mc{P}$, called the \emph{direction} \emph{field} of the observable, such that $d\mk{F}^0 = \bo{f} \lrcorner \bo{g}$. 
For each eigenvector $\bo{\psi}$ of $\widehat{\bo{F}}$, its presence mode $\psi$ is called an \emph{eigenmode} of the observable $\mk{F}$, and $\mk{F}(\psi)$ - an \emph{eigenvalue} of $\mk{F}$. The possible world $\mc{W}_{\psi}$ is called the \emph{eigenworld} of $\mk{F}$. \end{ax}
\begin{ax}[Dynamics] Left alone, the system evolves according to $d\mk{H}^0 = \bo{h} \lrcorner \bo{g}$, along the direction field $\bo{h}$ of a preferred observable, $\mk{H}$, the \emph{hyper}-\emph{Hamiltonian} of the system, such that the following diagram commutes for each possible world $\mc{W}$: 
\begin{displaymath} \begin{CD} \mc{W} @> \sigma >> \mc{H} \\ @V j VV @VV \tau V \\ \mc{P} @> \mk{H}^0 >> \R \end{CD} \quad , \end{displaymath} 
where $j$ is the inclusion map, and  $\sigma : \mc{W} \to \mc{H}$ is the fiber diffeomorphism.  \end{ax} 
\begin{ax}[Measurement] A \emph{measurement} of an observable $\mk{F}$ performed on a system is an assignment to the system of a hypersurface in an eigenworld of $\mk{F}$. A \emph{result} of a measurement of an observable $\mk{F}$ is an ordered triple $(\phi, \bo{a}, \delta)$, where $\phi$ and $\bo{a}$ are an eigenmode and the corresponding eigenvalue of $\mk{F}$, respectively; $\delta$ is a hypersurface in $\mc{W}_{\phi}$, called the \emph{carrier} of the system, defined by  
\begin{displaymath} \delta = \{\psi \in  \mc{W}_{\phi} \mid \tau^{\prime}(\psi) = t\} ,  \end{displaymath} where $t$ is the real part of $\bo{a}$ and $\tau^{\prime}$ is the pullback of $\tau$ under the fiber diffeomorphism $\sigma : \mc{W}_{\phi} \to \mc{H}$. The eigenvalue $\bo{a}$ can be given as an ordered pair $(t, \bo{x})$ where $t$ is interpreted as the \emph{time} \emph{of} \emph{realization} of the system in the eigenworld $\mc{W}_{\phi}$, and $\bo{x}$ is a pure imaginary quaternion, interpreted as the \emph{value} of the measurement. When a canonical basis $\bo{i}_{\beta}$ is specified, the value of the measurement is given by an ordered triple of real numbers. The geometry of $\delta$ is  defined by a metric $\bo{\mf{g}}^{\delta}$, the restriction of the metric $\bo{\mf{g}}^{\mc{W}_{\phi}}$ of the eigenworld $\mc{W}_{\phi}$ defined by the pullback of the proper metric $\bo{\mf{g}}^{\mc{H}}$ under the fiber diffeomorphism $\sigma$. \end{ax} 
\begin{ax}[Probability] If the system is in a state $\bo{\phi}$, then the probability $P$ that a measurement of $\mk{F}$ will return a result $(\psi, \bo{a}, \delta)$ is given by 
\begin{displaymath} P = \frac{\langle \bo{\phi} \mid \bo{\psi} \rangle \langle \bo{\psi} \mid \bo{\phi} \rangle} {\langle \bo{\phi} \mid \bo{\phi} \rangle \langle \bo{\psi} \mid \bo{\psi} \rangle} . \end{displaymath} \end{ax}
\section{Standard QM.}  Due to rather simple relationship between eigenvalues of Hermitian and anti-Hermitian operators in complex QM we can take either kind to represent observables. If we take anti-Hermitian operators, the Schr\"{o}dinger equation is 
$\bo{\dot{\phi}} = -\widehat{\bo{H}}(\bo{\phi})$, where $\widehat{\bo{H}}$ is a preferred anti-Hermitian operator, the \emph{Hamiltonian}.
Applying the procedure described in Section \ref{KINEMATICS} to a complex Hilbert space $(V_{\C}, \langle\cdot\mid\cdot\rangle)$, we get a Riemannian metric $\bo{g}$, a symplectic form $\tilde{\bo{\omega}}$ and a complex structure $\mc{I}$ on
$\mc{S}_{V_{\C}}$. Using the canonical identification \eqref{VM} we associate to each observable $\widehat{\bo{F}}$ an operator $\bo{F} := -\widehat{\bo{F}}$ on $S_{V_{\C}}$, vector field 
$\bo{f}$  and a real valued function $\mk{F}^1_{\C}$ on $\mc{S}_{V_{\C}}$,  such that $\mk{F}^1_{\C}\bo{i}$ is the expectation of $\widehat{\bo{F}}$. Then (see, e.~g. \cite{Sch96}), 
$d\mk{F}^1_{\C} = \bo{f} \lrcorner \tilde{\bo{\omega}}$.  
In particular,  Schr\"{o}dinger evolution is Hamiltonian evolution. Defining a real function $\mk{F}^0_{\C}$, such that $d\mk{F}^0_{\C}(\bo{f}^{\prime}) = d\mk{F}^1_{\C}(\mc{I}(\bo{f}^{\prime})), \forall  \bo{f}^{\prime}$, we can rewrite this as $d\mk{F}^0_{\C} = \bo{f} \lrcorner \bo{g}$. 
Given a canonical basis $\bo{i}_{\beta}$ on $\HH$, we embed $V_{\C}$ in a quaternionic Hilbert space $V$, identifying $\bo{i}$ with $\bo{i}_1$, and construct an observable $\mk{F} : \mc{S}_V \to \mc{S}_{\HH}$ with components $\mk{F}^{\beta}$ in the basis $\bo{i}_{\beta}$, such that $\mk{F}^0_{\C}$ and $\mk{F}^1_{\C}$ are restriction to $\mc{S}_{V_{\C}}$ of $\mk{F}^0$ and $\mk{F}^1$, respectively, and $\mk{F}^2, \mk{F}^3$ are two arbitrary constant real functions on $\mc{S}_V$. Thus standard QM can be embedded in HHQM. The geometric capacity of standard QM on its own, however, is insufficient to incorporate description of gravity for most systems of practical importance, because two (out of four) dimensions are collapsed in each possible world, and all the carriers are one dimensional. 


\begin{thebibliography}{99}
\bibitem{Ad95} S. L. Adler, \emph{Quaternionic Quantum Mechanics and Quantum 
Fields}, Oxford University Press, Oxford, UK (1995). 
\bibitem{CL00} J. Chen, J. Li, \emph{Quaternionic maps between hyperK\"{a}hler 
manifolds}, J. Diff. Geom. 55 (2000) 355-384. 
\bibitem{GM02} G. Gaeta, P. Morando, \emph{Hyper-Hamiltonian Dynamics}, J. Phys. 
\textbf{A35} (2002) 3925-3943, [math-ph/0204019]. 
\bibitem{MT02} P. Morando, M. Tarallo, \emph{Quaternionic Hamilton equations}, 
(2002) [math-ph/0204021]. 
\bibitem{Pos82} {\font \t = wncyr10 scaled 1200 \t M. M. Postnikov, Gruppy i algebry Li, 
Nauka, Moskva (1982)}. 
\bibitem{Sch96} T. A. Schilling, \emph{Geometry of Quantum Mechanics}, PhD Thesis, 
Pensylvania State University (1996). 
\bibitem{Tri95} V. Trifonov, \emph{A linear solution of the four-dimensionality problem}, Europhys. Lett., 32 (8), 1995, pp. 621-626, physics/0301044.
\bibitem{Tri03} V. Trifonov, \emph{Geometry} \emph{of} \emph{the} \emph{group} \emph{of} \emph{nonzero} \emph{quaternions}, phy-sics/0301052.
\end{thebibliography}
\end{document}